\documentclass[twocolumn,aps,pra,superscriptaddress,showpacs,amsmath]{revtex4-1}

\usepackage{braket}
\usepackage[per=slash,
            emulate=units, 
            range-phrase = -,
            range-units = single,
            exponent-product = \cdot,
            inter-unit-product = \cdot,
            separate-uncertainty = true]{siunitx}

\usepackage{hyperref}
\usepackage{graphicx}

\makeatother

\begin{document}

\title{Observation of the motional Stark shift in low magnetic fields}

\author{Manuel Kaiser}
\email[]{Manuel.Kaiser@student.uni-tuebingen.de}

\affiliation{Center for Quantum Science, Physikalisches Institut, Eberhard-Karls-Universit\"at
T\"ubingen, Auf der Morgenstelle 14, D-72076 T\"ubingen, Germany}

\author{Jens Grimmel}

\affiliation{Center for Quantum Science, Physikalisches Institut, Eberhard-Karls-Universit\"at
T\"ubingen, Auf der Morgenstelle 14, D-72076 T\"ubingen, Germany}

\author{Lara Torralbo-Campo}

\affiliation{Center for Quantum Science, Physikalisches Institut, Eberhard-Karls-Universit\"at
T\"ubingen, Auf der Morgenstelle 14, D-72076 T\"ubingen, Germany}

\author{Markus Mack}

\affiliation{Center for Quantum Science, Physikalisches Institut, Eberhard-Karls-Universit\"at
T\"ubingen, Auf der Morgenstelle 14, D-72076 T\"ubingen, Germany}

\author{Florian Karlewski}

\affiliation{Center for Quantum Science, Physikalisches Institut, Eberhard-Karls-Universit\"at
T\"ubingen, Auf der Morgenstelle 14, D-72076 T\"ubingen, Germany}

\author{Florian Jessen}

\affiliation{Center for Quantum Science, Physikalisches Institut, Eberhard-Karls-Universit\"at
	T\"ubingen, Auf der Morgenstelle 14, D-72076 T\"ubingen, Germany}

\author{Nils Schopohl}
\email[]{nils.schopohl@uni-tuebingen.de}

\affiliation{Center for Quantum Science, Institut f\"ur Theoretische Physik, Eberhard-Karls-Universit\"at
T\"ubingen, Auf der Morgenstelle 14, D-72076 T\"ubingen, Germany}

\author{J\'ozsef Fort\'agh}
\email[]{fortagh@uni-tuebingen.de}

\affiliation{Center for Quantum Science, Physikalisches Institut, Eberhard-Karls-Universit\"at
T\"ubingen, Auf der Morgenstelle 14, D-72076 T\"ubingen, Germany}

\date{\today}
\begin{abstract}
	We report on the observation of the motional Stark effect of highly excited $^{87}$Rb Rydberg atoms moving in the presence of a weak homogeneous magnetic field in a vapor cell. Employing electromagnetically induced transparency for spectroscopy of an atomic vapor, we observe the velocity-, quantum state- and magnetic field-dependent transition frequencies between the ground and Rydberg excited states. For atoms moving at velocities around $\SI{400}{\metre\per\second}$, the principal quantum number $n=100$ of the valence electron, and a magnetic field of $B=\SI{100}{\gauss}$, we measure a motional Stark shift of $\SI{\sim10}{\mega\hertz}$. Our experimental results are supported by numerical calculations based on a diagonalization of the effective Hamiltonian governing the valence electron of $^{87}$Rb in the presence of crossed electric and magnetic fields. 
\end{abstract}


\maketitle

	The motional Stark effect (MSE) introduces a coupling between the electronic structure of electronically bound particles and their center-of-mass motion in an external field. This correlation pointed out in the seminal work of Lamb \cite{lamb1952fine} plays an important role in fusion plasma diagnostics \cite{levinton1989magnetic,levinton1999motional} for measuring the magnetic fields, in astrophysics for the evaluation of hydrogen spectra in the vicinity of neutron stars \cite{pavlov1993finite,mori2002atomic}, as well as in solids for the magneto-Stark effect of excitons \cite{thomas1961magneto}. Although the atomic motion in magnetic fields is always accompanied by the MSE \cite{PhysRevLett.39.874,PhysRevA.18.1464,crosswhite1979motional,clark1984effects,0953-4075-28-17-003} and the center-of-mass motion of atoms becomes entangled with the internal dynamics \cite{avron1978separation,herold1981two,Schmelcher1997}, the MSE has received little attention so far.
	With advanced spectroscopic techniques \cite{Mohapatra.2007,arnoult2010optical} and the quest for the development of quantum devices based on hot atomic vapors \cite{julsgaard2004experimental,appel2008quantum,low2009magneto,cho2010atomic,kubler2010coherent,Fang2015}, the MSE of atoms becomes a measurable quantity and adds features of key importance: atoms are no longer described by a single wave function but a two-body core-electron wave function that is coupled through a pseudomomentum. At the same time, atoms are highly controllable quantum systems and enable the development of general models and experimental test opportunities for the coupled two-body problem of charged particles in external fields with direct impact on research on plasmas, electron-hole pairs \cite{gor1968contribution,PhysRevB.95.245205}, and particle-antiparticle symmetries \cite{PhysRevD.88.105017}.
		
	In our paper we extend the investigation of the MSE to low magnetic fields and quantify it on an element other than hydrogen. For $^{87}$Rb Rydberg atoms we measured spectral shifts up to $\SI{10}{\mega\hertz}$ with a spectroscopic resolution of $\SI{2}{\mega\hertz}$ for the principal quantum number $n=100$ and a field of $\SI{100}{\gauss}$, using the phenomenon of electromagnetically induced transparency (EIT) on atoms in a thermal vapor cell. We complement the experimental data with numerical calculations of an atom in crossed magnetic and electric fields and thereby show that our theory based on an effective two-body system describes the complex rubidium Rydberg atom well.
	
	\begin{figure}
		\includegraphics{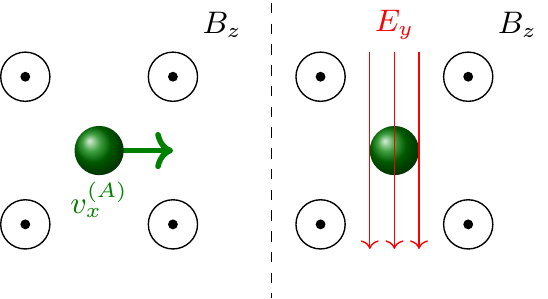}
		\caption{\label{fig:Lorentz_field} Atoms moving in the laboratory frame at velocity $\mathbf{v}^{(\text{A})}$ in the presence of a magnetic induction field $\mathbf{B}$ are in their reference frame subject to a Lorentz electric field [Eq. \eqref{eq:Lorentz electrical field}].}
	\end{figure}	
	
	The elementary attributes of atoms that govern all interaction processes with the electromagnetic field are charge and spin. Pieced together from orbital angular momentum and spin, the magnetic moments of atoms interact with the field of magnetic induction $\mathbf{B}$ and give rise to various splittings and changes of the internal atomic energy structure. As a consequence, the spectrum of atoms moving in the presence of a $\mathbf{B}$ field may, besides the Doppler shift, be altered because a charge moving at velocity $\mathbf{v}$ in the presence of a magnetic induction field experiences in its (instantaneous) rest frame a Lorentz electric field,
	\begin{equation}
		\mathbf{E}_{L}=\mathbf{v}\times\mathbf{B}.
		\label{eq:Lorentz electrical field}
	\end{equation}
	This causes the positively charged nucleus and the electrons of an atom to sense a Lorentz force acting in opposite directions, when moving in a magnetic field (see Fig. \ref{fig:Lorentz_field}). Consequently excited atoms in motion will emit a spectrum featuring not only the usual Doppler shift but also a Stark effect whose magnitude is primarily dependent on the atom's velocity and flight direction.
	
	In distinction from the hydrogen atom (and its isotopes) the theoretical description of the electronic structure of heavier atoms poses a formidable many-body problem that cannot be solved exactly. Therefore, one has to rely on an approximate description in terms of an effective hydrogenlike problem, in which the bound-state spectrum of the excited valence electron of an alkali-metal atom with mass $M$ can be well described by the spherically symmetric effective potential $V_{\text{eff}}(r)$ of Marinescu \textit{et al.} \cite{marinescu1994dispersion}. Here the variable $r=|\mathbf{r}^{(e)}-\mathbf{r}^{(c)}|$ denotes the distance between the valence electron at position $\mathbf{r}^{(e)}$ and a collective coordinate $\mathbf{r}^{(c)}$ that determines the position of the center-of-mass of the ionic core with charge $Z|e|-(Z-1)|e|=|e|$ and mass $m^{(c)}=M-m^{(e)}$. 
	
	We therefore propose to describe the spectrum of an alkali-metal Rydberg atom moving in the presence of external electromagnetic fields with the effective two-body Hamiltonian:
	\begin{align}
		\label{eq:two-body_Hamiltonian}
		H^{(2)}=&\frac{[\mathbf{p}^{(c)}-|e|\mathbf{A}(\mathbf{r}^{(c)})]^{2}}{2m^{(c)}}+\frac{[\mathbf{p}^{(e)}+|e|\mathbf{A}(\mathbf{r}^{(e)})]^{2}}{2m^{(e)}}\\&+V_{\text{eff}}(|\mathbf{r}^{(e)}-\mathbf{r}^{(c)}|)+\frac{\mu_{B}}{\hbar}g_{S}\mathbf{S}\cdot\mathbf{B}\nonumber\\
		&+\left|e\right|\boldsymbol{\mathcal{E}}\cdot(\mathbf{r}^{(e)}-\mathbf{r}^{(c)}).\nonumber
	\end{align}
	Here $\boldsymbol{\mathcal{E}}$ is a homogeneous static external electric field, and $\mathbf{B}=\textrm{rot}\mathbf{A}$ is a homogeneous external magnetic induction field, in the symmetric gauge $\mathbf{A}(\mathbf{x})=\frac{1}{2}\mathbf{B}\times\mathbf{x}$. It is convenient to rewrite $H^{(2)}$ in the center-of-mass frame with new variables, $\mathbf{R}=\frac{m^{(e)}\mathbf{r}^{(e)}+m^{(c)}\mathbf{r}^{(c)}}{M}$ and $\mathbf{r}=\mathbf{r}^{(e)}-\mathbf{r}^{(c)}$ with the conjugate momenta $\mathbf{P}=\mathbf{p}^{\left(c\right)}+\mathbf{p}^{\left(e\right)}=\frac{\hbar}{i}\boldsymbol{\nabla}_{R}$ and $\mathbf{p}=\frac{m^{(c)}}{M}\mathbf{p}^{(e)}-\frac{m^{(e)}}{M}\mathbf{p}^{(c)}=\frac{\hbar}{i}\boldsymbol{\nabla}_{r}$. However, the associated Schr\"odinger eigenvalue problem for this Hamiltonian is not separable, because for $\mathbf{B\neq0}$ the total momentum $\mathbf{P}$ is not conserved. Instead the Cartesian components $\mathcal{P}_{a}$ of the pseudomomentum
	\begin{align}
		\label{eq:pseudo-momentum}
		\boldsymbol{\mathcal{P}} &=\mathbf{p}^{(c)} +|e|\mathbf{A} \left(\mathbf{r}^{(c)}\right) +\mathbf{p}^{(e)}-|e|\mathbf{A}\left(\mathbf{r}^{(e)}\right)\\
		 &= \mathbf{P}-\frac{|e|}{2}\mathbf{B}\times\mathbf{r}\nonumber 
	\end{align}
	are conserved \cite{johnson1983interaction}:
	\begin{equation}
		\label{eq:commutator relations}
		\left[\mathcal{P}_{a},H^{\left(2\right)}\right] = 0,\qquad
		\left[\mathcal{P}_{a},\mathcal{P}_{a'}\right] = 0.
	\end{equation}
	These commutator relations engender the existence of a complete system of orthonormal two-body eigenfunctions $\varPsi_{\mathbf{k},\nu}(\mathbf{r}^{(e)},\mathbf{r}^{(c)})\equiv\widetilde{\varPsi}_{\mathbf{k},\nu}(\mathbf{R},\mathbf{r})$ that are eigenfunctions of both operators, $\mathcal{P}_{a}$ and $H^{(2)}$, simultaneously:
	\begin{align}
		\label{eq: two-body eigenvalue problem}
		H^{(2)}\widetilde{\varPsi}_{\mathbf{k},\nu} &= E_{\mathbf{k},\nu}\widetilde{\varPsi}_{\mathbf{k},\nu}\\
		\mathcal{P}_{a}\widetilde{\varPsi}_{\mathbf{k},\nu} & = \hbar k_{a}\widetilde{\varPsi}_{\mathbf{k},\nu}.\nonumber
	\end{align}
	Here $\nu$ is a multi-index labeling intrinsic quantum states of the valence electron. It follows, assuming box normalization
	with regard to the center-of-mass variable $\mathbf{R}$, that the sought eigenfunctions of $H^{(2)}$ and $\mathcal{P}_{a}$ are \cite{gor1968contribution}
	\begin{equation}
		\widetilde{\varPsi}_{\mathbf{k},\nu}(\mathbf{R},\mathbf{r}) =\frac{\exp\left[i\left(\mathbf{k}+\frac{|e|}{2\hbar}\mathbf{B}\times\mathbf{r}\right)\cdot\mathbf{R}\right]}{\sqrt{L^{3}}}\psi_{\mathbf{k},\nu}\left(\mathbf{r}\right),
	\end{equation}
	where $\psi_{\mathbf{k},\nu}(\mathbf{r})$ is an eigenfunction associated with a single-particle Hamiltonian $H_{\mathbf{k}}^{(1)}$ depending parametrically on the eigenvalue $\hbar\mathbf{k}$ of the pseudomomentum $\boldsymbol{\mathcal{P}}$ \cite{avron1978separation}:
	\begin{equation}
		\label{eq:single_particle_Hamiltonian_I}
		H_{\mathbf{k}}^{(1)}\psi_{\mathbf{k},\nu}(\mathbf{r})=E_{\mathbf{k},\nu}\psi_{\mathbf{k},\nu}(\mathbf{r}).
	\end{equation} 
	We then find that Eq. \eqref{eq:single_particle_Hamiltonian_I} has, besides the terms dependent on $\hbar\mathbf{k}$, the guise of the standard Hamiltonian of the valence electron of an alkali-metal atom \cite{gallagher}, including paramagnetic, diamagnetic, and electric-field interactions:
	\begin{align}
		\label{eq:single particle Hamiltonian II}
		H_{\mathbf{k}}^{(1)} =&\frac{\hbar^{2}\mathbf{k}^{2}}{2M}+\frac{\mathbf{p}^{2}}{2\mu}+V_{\text{eff}}(r)+\frac{\mu_{B}}{\hbar}\left(g_L\mathbf{L}+g_{S}\mathbf{S}\right)\cdot\mathbf{B}\nonumber\\ &+\left|e\right|\left(\boldsymbol{\mathcal{E}}+\frac{\hbar\mathbf{k}}{M}\times\mathbf{B}\right)\cdot\mathbf{r}+\frac{\left|e\right|^{2}}{8\mu}\left(\mathbf{B}\times\mathbf{r}\right)^{2},
	\end{align}
	with effective mass $\frac{1}{\mu}=\frac{1}{m^{\left(e\right)}}+\frac{1}{m^{\left(c\right)}}$, $g$-factor $g_L = 1-\frac{m^{(e)}}{m^{(c)}}$, and orbital angular momentum operator $\mathbf{L}=\mathbf{r}\times\mathbf{p}$.
	For the atom velocity in the Heisenberg picture one obtains $\mathbf{v}^{(A)}=\frac{d}{dt}\mathbf{R}=\frac{1}{i\hbar}[\mathbf{R},H^{\left(2\right)}]=\frac{1}{M}\left(\mathbf{P}+\frac{\left|e\right|}{2}\mathbf{B}\times\mathbf{r}\right)$. We can now eliminate the center-of-mass momentum $\mathbf{P}$ instead of the pseudomomentum $\boldsymbol{\mathcal{P}}$, see Eq. \eqref{eq:pseudo-momentum}, and obtain
	\begin{equation}
		\label{eq:velocity pseudomomentum}
		\mathbf{v}^{(A)} = \frac{1}{M}(\boldsymbol{\mathcal{P}} + |e|\mathbf{B}\times\mathbf{r}).
	\end{equation}
	\begin{figure*}
		\includegraphics{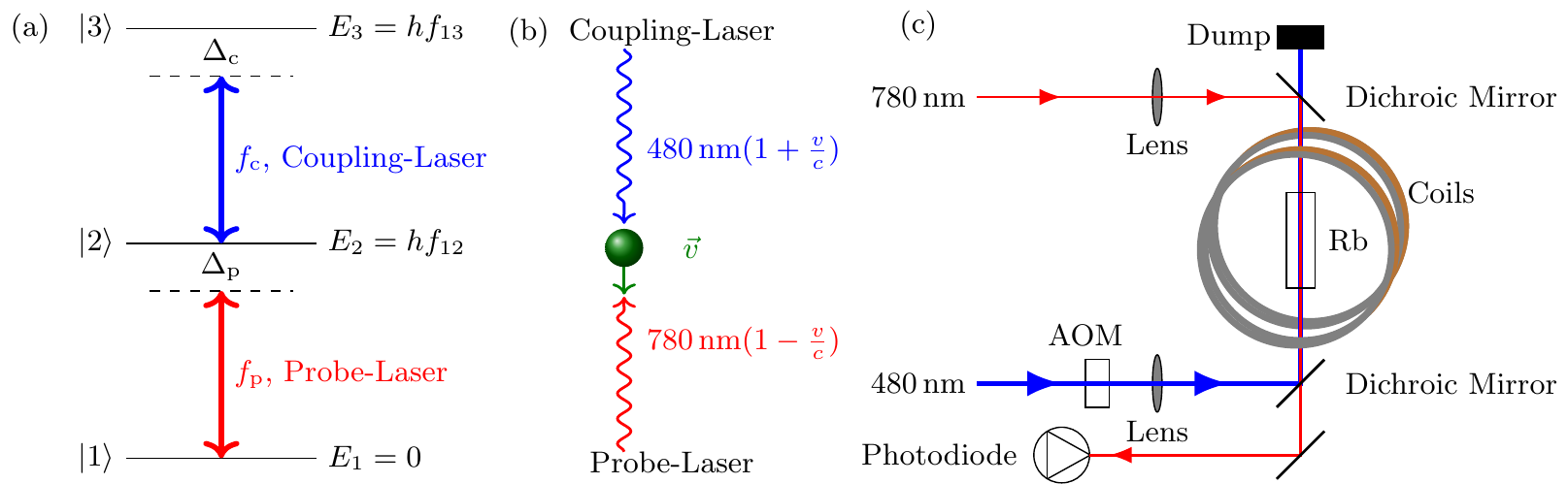}
		\caption{\label{fig:EIT-scheme} (a) Energy-level diagram for EIT spectroscopy in ladder configuration. A strong coupling beam at $\SI{\sim480}{\nano\metre}$ induces a narrow transparency window for a weak probe beam at $\SI{\sim780}{\nano\metre}$. (b) Detunings from the resonance frequencies can be compensated by Doppler shifts of moving atoms. By detuning both lasers reversely only a selected velocity class contributes to the EIT signal. \label{fig:Osetup}(c) Optical setup for the EIT spectroscopy. The coupling laser is intensity modulated with an  acousto-optic modulator (AOM) and focused inside a vapor cell. It is overlapped in the cell with the counterpropagating probe laser, whose transmission is detected on a photodiode. The cell at $\SI{\sim50}{\celsius}$ is placed in between two magnetic coils in Helmholtz configuration.}
	\end{figure*}
	For strong magnetic fields the term $|e|\mathbf{B}\times\mathbf{r}$ can have a high impact on the atomic motion \cite{pohl2009cold}. However, in weak magnetic fields such as considered here and at thermal atom speeds $\mathbf{v}^{\left(A\right)}$ the term can be neglected on the level of accuracy of our measurements up to Rydberg levels $n<110$. This permits replacing $\mathbf{v}^{(A)}\leftarrow\frac{\hbar\mathbf{k}}{M}$ and interpreting the term $\frac{\hbar\mathbf{k}}{M}\times\mathbf{B}$ in the effective single-particle Hamiltonian Eq. \eqref{eq:single particle Hamiltonian II} as a Lorentz electric field; see Eq. \eqref{eq:Lorentz electrical field}. For Rydberg levels as high as $n=150$ the correction to $\mathbf{v}^{(\text{A})}$ due to the dipole term in Eq. \eqref{eq:velocity pseudomomentum} amounts to $\SI{\sim100}{\meter\per\second}$. The difference between $\mathbf{v}^{(A)}$ and $\frac{\hbar\mathbf{k}}{M}$ may be seen better in other experiments, for example by monitoring the dipole mode of an ultracold alkali-metal atom cloud moving in a magnetic trap, by separating an atomic beam in a Stern-Gerlach-like experiment by laser excitation and thereby changing the internal energy structure or by measuring the structure factors (quantum correlations) of a classical gas during excitation to Rydberg states.
	
	Even though the MSE is similar to the regular Stark effect at first sight, there is an important difference, as a $\mathbf{B}$ field cannot do work on a moving atom and therefore cannot ionize it. Hence, using Eq. \eqref{eq:Lorentz electrical field}, we can still analyze the MSE numerically on the basis of Eq. \eqref{eq:single particle Hamiltonian II} as if it was a system in crossed fields configuration. The position operator $\mathbf{r}$ can be expressed in spherical coordinates where the angular parts can be evaluated with matrix elements from \cite{bethe2012quantum}. For the calculation of the radial wave functions we use the parametric model potential $V_{\text{eff}}(r)$ from \cite{marinescu1994dispersion}, adapted to the experimental situation with the theory of \cite{sanayei2015quasiclassical}. We then calculate the energy levels of the crossed fields system using an energy matrix diagonalization similar to \cite{Zimmerman:1979}. The energy levels in zero field are calculated using quantum defects from \cite{mack2011measurement}.
	For each energy eigenvalue $E_{\mathbf{k},\nu}$ we represent the corresponding eigenvector of $H_{\mathbf{k}}^{(1)}$ as a linear combination of zero-field eigenstates, to calculate the dipole transition strength taking into account the laser polarizations as in \cite{grimmel2015measurement}. These eigenvectors for states in external fields are also used to estimate the dipole moment $\mathbf{d}_{\mathbf{k},\nu} = -|e|\Braket{\Psi_{\mathbf{k},\nu}|\mathbf{r}|\Psi_{\mathbf{k},\nu}}$ from Eq. \eqref{eq:velocity pseudomomentum}, resulting in a calculated difference of velocity $\mathbf{v}^{(A)}$ and the pseudomomentum on the order of $\SI{0.1}{\meter\per \second}$ for the conditions of our experiment.

	On the experimental side, we analyze the motional Stark shifts by using a two-photon spectroscopy method based on EIT in a ladder scheme similar to \cite{Mohapatra.2007}. A strong laser which couples the intermediate state $5P_{3/2}$ and a Rydberg state leads to a narrow transparency window for a laser probing the lower $5S_{1/2}\to5P_{3/2}$ transition, in case both lasers are in resonance with an atomic transition [see Fig. \ref{fig:EIT-scheme}(a)]. The difference in frequency of the two transitions allows us to select a velocity class $v^{(A)}$ by detuning the laser frequencies $f_{p}$ and $f_{c}$ according to the Doppler shifted two-photon resonance condition
	\begin{equation}
		\Delta_{p}+\Delta_{c}= v^{\text{(A)}} \left(\frac{f_{p}-f_{c}}{c}\right),
	\end{equation}
	with the detunings $\Delta_{p}$ and $\Delta_{c}$ of the probe and coupling laser, respectively, and the speed of light $c$ [see Fig. \ref{fig:EIT-scheme}(b)]. We can select atoms at rest ($\SI{100}{\meter\per \second}$) from a vapor with Maxwell-Boltzmann distributed atom velocities by fixing the probe laser frequency to the atomic transition, i.e., $\Delta_{p}=\SI{0}{\mega\hertz}$ ($\Delta_p = f_p\frac{v^{\text{(A)}}}{c}\SI{\approx 128}{\mega\hertz}$). If we scan the coupling laser over the atomic resonance, the maximum transparency in zero field then appears for a coupling laser detuning of $\Delta_c = \SI{0}{\mega\hertz}$ ($\Delta_c \SI{\approx - 209}{\mega\hertz}$).
	
	For our measurement we use a standard rubidium vapor cell with a length	of $\SI{75}{\milli\metre}$ at $\SI{\sim50}{\celsius}$ enabling us to obtain spectra from a large range of velocity classes up to $\SI{\sim600}{\metre\per\second}$. The cell is placed in between a pair of coils in Helmholtz configuration which provides fields up to $\SI{100}{\gauss}$ [see Fig. \ref{fig:Osetup}(c)]. The magnetic field is calibrated using a Hall sensor with an error smaller than $\SI{0.1}{\gauss}$ leaving only a small offset magnetic field. Stray electric fields are effectively canceled by charges inside the cell \cite{Mohapatra.2007}.
	\begin{figure}
		\includegraphics{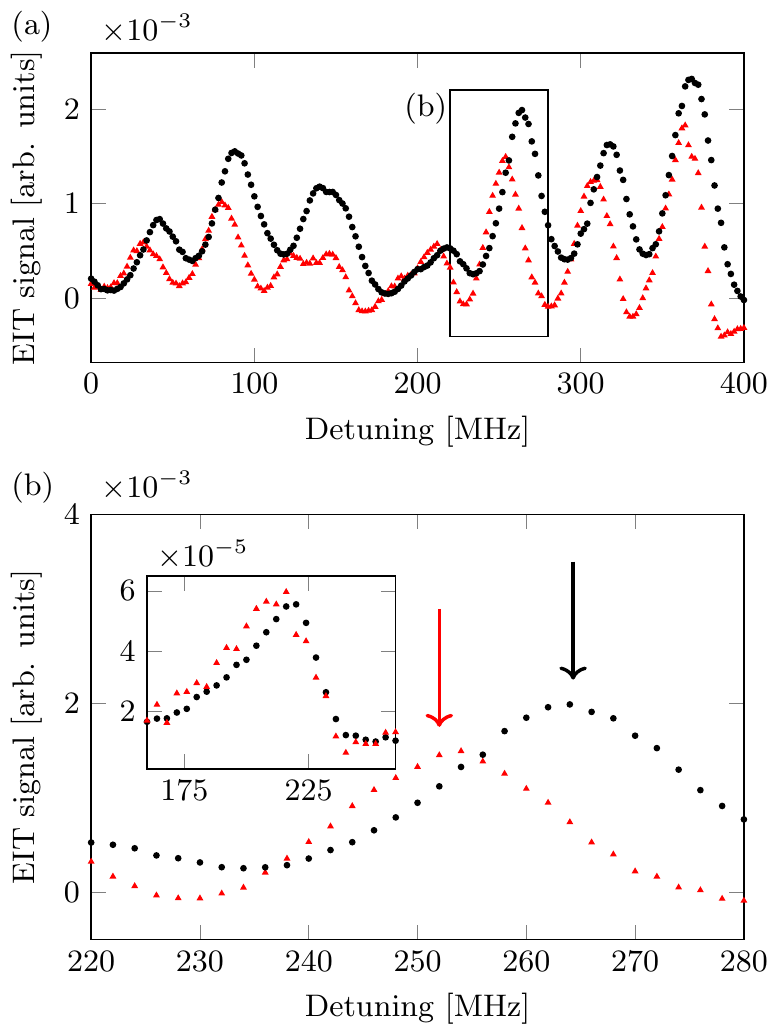}
		\caption{\label{fig:100D_multi_1} (a) The EIT signal for the $100\text{D}_{5/2}$ state is shown in dependence on the coupling laser detuning $\Delta_c$ relative to the selected velocity class. The motional Stark shifts are on the order of $\SI{10}{\mega\hertz}$ in a field of $\SI{98}{\gauss}$ for atoms at rest (black dots) compared to atoms moving at $v^{(A)}=\SI{390}{\metre\per\second}$ (red triangles).
			(b) EIT signal in dependence of $\Delta_c$ for the $100\text{D}_{5/2}$ in detail for two velocity classes $v^{(A)}=\SI{0}{\metre\per\second}$ (black dots) and $v^{(A)}=\SI{390}{\metre\per\second}$ (red triangles) with the corresponding calculated resonances (arrows). The MSE vanishes for atoms moving parallel to the $\mathbf{B}$ field (inset).}
	\end{figure}
	
	The linearly polarized coupling laser (TA-SHG pro, Toptica) at $\SI{480}{\nano\metre}$ with a power of $\SI{\sim80}{\milli\watt}$ is focused inside the cell ($\SI{\sim150}{\micro\metre}$ $\frac{1}{e^2}$ width). An also linearly polarized but counterpropagating probe beam (DL pro, Toptica) at $\SI{780}{\nano\metre}$ is overlapped with the coupling laser in the cell and is detected with a photodiode (APD110A, Thorlabs). For a better signal-to-noise ratio we use a lock-in amplifier (HF2LI, Zurich Instruments) which modulates the intensity of the coupling laser with an AOM and demodulates the probe laser signal from the photodiode. Each of the lasers is locked to a Fabry-Perot interferometer (FPI 100, Toptica). The FPI of the probe laser is locked to a frequency comb (FC 1500, Menlo Systems). The coupling laser FPI is controlled by a wavelength meter (WS Ultimate 2, HighFinesse) which is calibrated to the beat of the coupling laser frequency at $\SI{960}{\nano\metre}$ with the frequency comb. Within the measurement times the frequency accuracy of our laser system is better than $\SI{2}{\mega\hertz}$.
	
	We investigate the MSE by comparing the shifts at different velocity classes in a magnetic field. The probe beam is always on resonance with the corresponding Doppler shifted transition frequency. The coupling laser is scanned and at each step the photodiode signal is recorded for $\SI{10}{\second}$. The $\mathbf{B}$ field is set to a fixed value for each cycle. We estimate the errors of the peak-center frequencies by fitting Lorentzian peaks to the obtained EIT spectrum, averaging over multiple measurement cycles and adding the uncertainties of $\SI{2}{\mega\hertz}$ of the lasers. The measured spectra are fitted to the numerical calculations with a fixed offset magnetic field for all velocity classes as the only free parameter.
	\begin{figure}
		\includegraphics{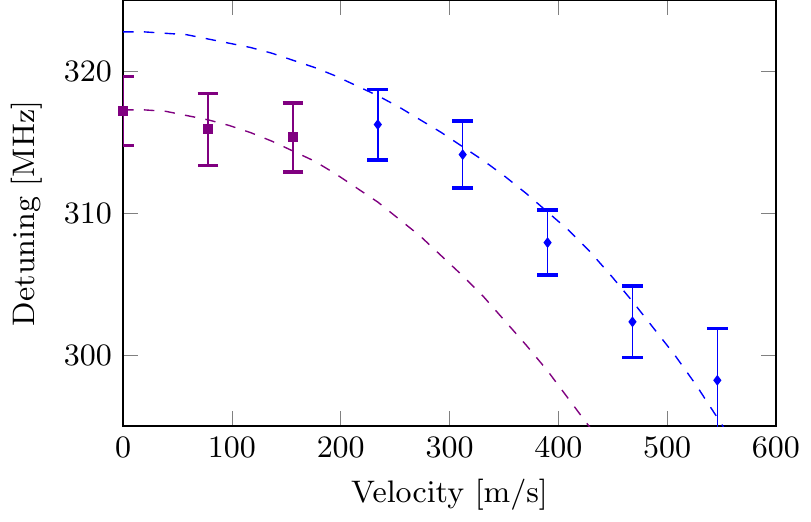}
		\caption{\label{fig:100D_vel} Velocity dependence of the MSE for the $100\text{D}_{5/2}$ state. The squares and diamonds with error bars are the measured transmission peaks for coupling laser detunings $\Delta_c$ representing the resonance frequencies in a field of $\SI{98}{\gauss}$. The two dashed lines are calculated energy levels. Mixing effects exchange oscillator strengths between the states and the measured transmission peak can be assigned to either of them.}
	\end{figure}
	
	The motional electric field for atoms moving with $\SI{\sim390}{\meter\per\second}$ in a field of $\SI{98}{\gauss}$ is $\SI{\sim0.038}{\volt\per\centi\meter}$. This results in a shift of $\SI{\sim10}{\mega\hertz}$ for the measured spectrum of the $100\text{D}_{5/2}$ state [see Fig. \ref{fig:100D_multi_1}(a)]. A single resonance is shown in detail [see Fig. \ref{fig:100D_multi_1}(b)] where the theory values (arrows) are calculated as described before with a matrix dimension of 20\,000, where a variation in the dimension only accounts for a submegahertz variation in frequency. Within the limits of our experimental accuracy we find good agreement between the experiment and the theory for an offset magnetic-field parameter smaller than $\SI{1}{\gauss}$. Moreover, they match well for measurements of other states (not shown here), which entails the demonstration of the strong dependence of the MSE on the quantum state.
	
	Furthermore atoms resting and moving parallel to the $\mathbf{B}$ field do not show a motional Stark shift [inset of Fig. \ref{fig:100D_multi_1}(b)]. For this measurement we changed the direction of the magnetic field and recorded EIT spectra of the $100\text{D}_{5/2}$ state in a field of $\SI{98}{\gauss}$. Due to geometrical restrictions a shorter cell was used for this part of the experiment. Even though no shift is observed, the transmission peak shows an asymmetry. Simulations of the line shape of the EIT signal taking into account the MSE for velocity components perpendicular to the optical axis indicated a much smaller asymmetry. We attribute this discrepancy to an additional inhomogeneity of the magnetic field.
	
	Beyond the dependence on the quantum state and the direction of $\mathbf{B}$ and $\mathbf{v}^{(\text{A})}$, the absolute value of the velocity component perpendicular to the field plays an important role. This velocity dependence of the shift is shown in Fig. \ref{fig:100D_vel}. The velocities correspond to probe laser detunings between 0 and $\SI{700}{\mega\hertz}$. From our numerical calculation we can assign the measured peak to two different substates whose intensities are transferred from one state to another through the MSE at around $\SI{250}{\meter\per \second}$. 
	
\begin{figure}
		\includegraphics{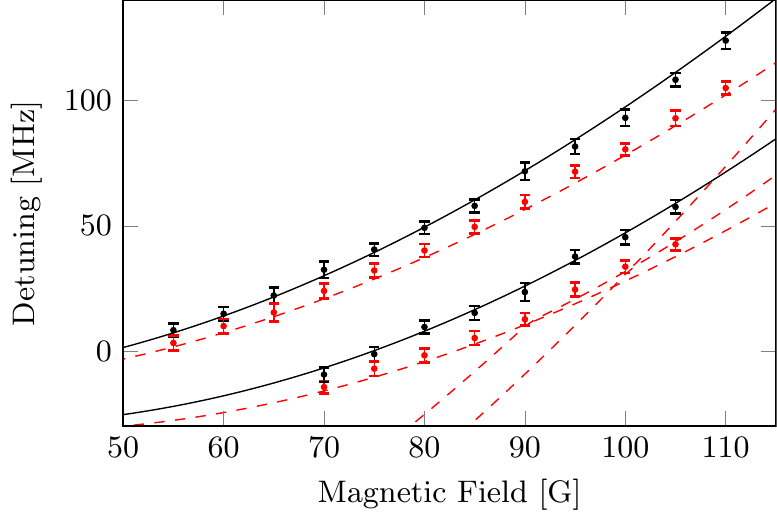}
		\caption{\label{fig:100D_two} The EIT resonances of the two velocity classes $v^{(A)}=\SI{0}{\metre\per\second}$ (black lines) and $v^{(A)}=\SI{390}{\metre\per\second}$ (dashed red lines) for the $100\text{D}_{5/2}$ state are shifted by the MSE dependent on the magnetic field. The detunings $\Delta_c$ relative to the selected velocity classes for the maximal probe laser transmission (dots with error bars) match the theory lines well. The oscillator strength for different states (red lines in the lower part) are altered through the MSE and the evaluated peak does not follow one theory line.}
	\end{figure}

	Furthermore the $\mathbf{v}^{(\text{A})}\times\mathbf{B}$ term relates the MSE to the magnetic field which is shown in Fig. \ref{fig:100D_two}. For magnetic fields lower than $\SI{50}{\gauss}$ the shift is smaller than the uncertainties from the laser system and therefore not shown here. For zero field the energy levels of the states coincide. The lower resonance lines indicate a transfer of oscillator strengths between different states through the appearance of anticrossings of $m$ states due to the motional electric field.
	
	In conclusion, our work expands the experimental investigations of the MSE to low static magnetic fields. We observed the motional Stark effect on $^{87}$Rb Rydberg atoms in a vapor cell using EIT spectroscopy with an accuracy better than $\SI{2}{\mega\hertz}$. At $\SI{100}{\gauss}$ the shifts are on the order of $\SI{10}{\mega\hertz}$ for the $100\text{D}_{5/2}$ state, which is in good agreement with the results of our numerical calculation based on an energy matrix diagonalization of the atom in crossed fields.
	We introduced a two-body model system for alkali-metal Rydberg atoms along with experimental data and conclude that it opens opportunities in describing many-body systems. The theoretical description of the MSE by a two-body Hamiltonian also confirms that the influence of the coupling of internal dynamics to the collective motion of the atom is small, but we estimate it to become crucial for states of $n\ge 150$ for a magnetic field of $\SI{100}{\gauss}$. Finally, calculations of atomic multielectron spectra in crossed fields configurations can be tested experimentally using the MSE as the condition $\boldsymbol{\mathcal{E}}\perp \mathbf{B}$ is exactly fulfilled with $\boldsymbol{\mathcal{E}} = \mathbf{E}_L$ which otherwise is hardly achievable in experiments with two external fields.
\appendix*

\begin{acknowledgments}
	This work was financially supported by  Deutsche Forschungsgemeinschaft through SPP 1929 (GiRyd). 
\end{acknowledgments}

\bibliography{literature}

\end{document}